\documentclass[pra,twocolumn]{revtex4-1}
\usepackage{graphicx,bm,float,color,hyperref,amssymb,amsmath,physics,cancel}
\hypersetup{colorlinks=true linkcolor={blue}}
\newcommand{\Ndot}{{\dot{N}}}

\def\<{{\langle}}
\def\>{{\rangle}}

\newcommand{\al}{\alpha}

\newcommand{\de}{\delta}
\newcommand{\D}{\Delta}
\newcommand{\e}{\epsilon}

\newcommand{\g}{\gamma}
\newcommand{\G}{\Gamma}
\newcommand{\kap}{\kappa}
\newcommand{\lm}{\lambda}

\newcommand{\w}{\omega}
\newcommand{\W}{\Omega}

\newcommand{\s}{\sigma}

\newcommand{\p}{\partial}

\newcommand{\mbk}{\mathbf{k}}

\newcommand{\mbr}{\mathbf{r}}

\newcommand{\bsphi}[1]{\boldsymbol{\phi}}

\newcommand{\mcal}[1]{\mathcal{#1}}

\newcommand{\til}[1]{\tilde{#1}}

\newcommand{\nn}{\nonumber\\}
\newcommand{\up}{\uparrow}
\newcommand{\down}{\downarrow}

\newcommand{\ben}{\begin{equation}}
\newcommand{\een}{\end{equation}}
\newcommand{\eq}[1]{Eq.~\eqref{#1}}
\newcommand{\fig}[1]{Fig.~\ref{#1}}

\newcommand{\onehalf}{\frac{1}{2}}

\def\simlt{\mathrel{\lower .3ex \rlap{$\sim$}\raise .5ex \hbox{$<$}}}
\def\simgt{\mathrel{\lower .3ex \rlap{$\sim$}\raise .5ex \hbox{$>$}}}

\begin{document}

\title{Quantum efficiency of a single microwave photon detector based on a semiconductor double quantum dot}
\author{Clement H. Wong}
\affiliation{Department of Physics, University of Maryland-College Park, College Park, Maryland 20742, USA}
\author{ Maxim G. Vavilov}
\affiliation{Department of Physics, University of Wisconsin-Madison, Madison, Wisconsin 53706, USA}
\date{\today}
\begin{abstract}
Motivated by recent interest in implementing circuit quantum electrodynamics with semiconducting quantum dots, we consider a double quantum dot (DQD) capacitively coupled to a superconducting resonator that is driven by the microwave field of a superconducting transmission line. 
We analyze the DQD current response using input-output theory and show that the resonator-coupled DQD  is a sensitive microwave single photon detector
Using currently available experimental parameters of DQD-resonator coupling and dissipation, including the effects of  $1/f$ charge noise and phonon noise,  we determine the parameter regime for which incident photons are completely absorbed and near unit $\gtrsim$ 98\% efficiency can be achieved.  We show that this regime can be reached by using very high quality resonators with {quality factor $Q\simeq 10^5$}.
\end{abstract}
\maketitle

\section{Introduction} 

High performance, single photon detectors are essential tools in quantum optics, with applications in optical quantum information processing, communication, cryptography, and metrology \cite{hadfieldNATP09}.
Single photon detection in the microwave regime has similar applications in the emerging field of microwave quantum photonics, made possible by recent advances in  implementing circuit quantum electrodynamics (cQED) with superconducting circuit technology \cite{nakamura12,nakamuraIEEE13}, but are more difficult to achieve because microwave photons have energy five orders of magnitude less than optical photons.   Besides photonics, microwave photon detectors have applications in astronomy and cosmology, for example, in  measuring the cosmic microwave background \cite{spieler04}.  {Microwave radiometers are commonly used in  meteorological and oceanographic remote-sensing.}

Recent theoretical proposals and experimental developments in microwave photon detectors are based on Josephson junctions \cite{peropadrePRA11,chenPRL11,romeroPRL09,koshinoPRL13,koshinoPRA15,poudelPRB12}.
At the same time, experimental progress in implementing cQED with semiconducting quantum dots is showing promise \cite{dengCMkondo15,dengNL15,dengPRL15,zhangAPL14,bassetPRB13,kulkarniPRB14,xuPRB13,*xuPRB13fc,peterssonNAT12,freyPRL12,jinPRB11,childressPRA04}.
Currently available resonator-quantum dot systems already allow for some interesting quantum optics applications such as on-chip single emitter masers \cite{liuPRL14,liuSCI15}, and tunable self-interaction and dissipation of the resonator photons induced by the quantum dot \cite{schiroPRB14,greentreeNATP06}. 
When a quantum dot is connected to electric leads, this system provides a platform for studying the interplay between quantum impurity physics and quantum optics \cite{lehurCM15}.

In this paper, we propose a photon detector based on photon assisted tunneling of electrons through a double quantum dot (DQD), and determine the quantum efficiency of single photon detection. We identify the parameter regime where reflection of input photons from the resonator vanishes, so that near unit efficiency can be achieved with currently available experimental parameters.
Such a high efficiency is possible even in the presence of strong DQD dissipation because the detection process takes advantage of fast dot-lead tunneling relative to the DQD inelastic decay rate, and does not require strong DQD-resonator coupling relative to the DQD dissipation rates.

The zero reflection regime is also relevant in the context of quantum computation, since it enables distant transmission of quantum information required in quantum cryptography and communication \cite{ciracPRL97,pinotsiPRL08}. We also note that when the input photons come from a hot thermal source, this device acts as a quantum heat engine~\cite{vavilovPRL06,bergenfeldtPRL14}.  

This paper is organized as follows.  In Sec.~\ref{sec:model}, we introduce our theoretical model for the transmission line carrying incoming photons and the photon detector. In Sec.~\ref{sec:EOM}, we present the equations of motion governing the system dynamics, and in Sec.~\ref{sec:SSS}, we present the steady state solution, which is used to derive an analytic expression for the quantum efficiency of photon detection in Sec.~\ref{sec:eff}.  In Sec.~\ref{sec:ref}, we discuss how the reflected signal is incorporated inour model, and in Sec.~\ref{sec:optimal}, we find the optimal parameter regime.  A brief summary of the decoherence model we used for the DQD is given in the appendix.


\begin{figure}[t]
\begin{center}
\includegraphics[width=\linewidth]{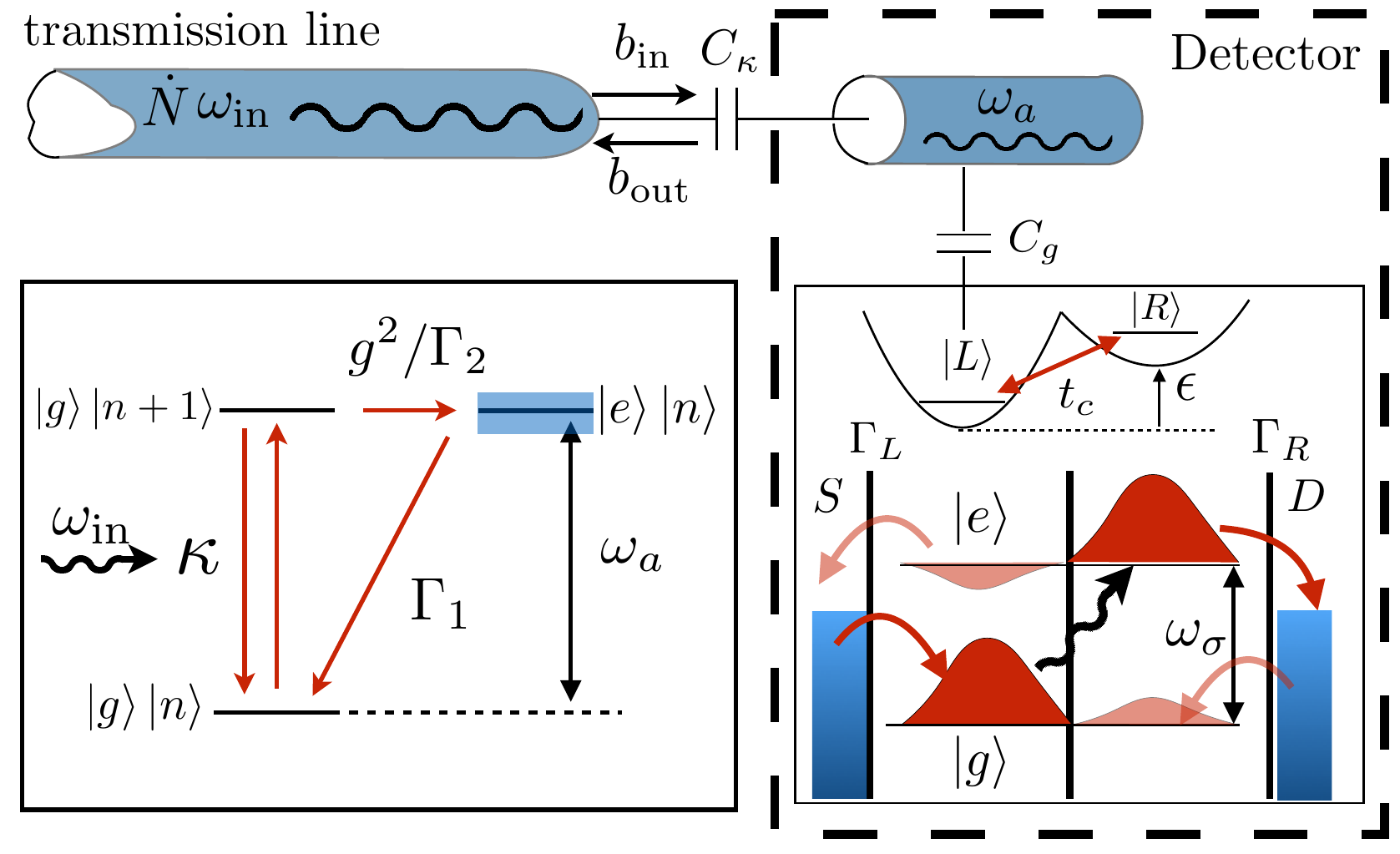}
\caption{(Color online)
Top left: A transmission line carries a continuous wave of microwave photons with flux $\dot{N}$ and frequency $\w_{\rm in}$.
Right dashed box:  The photon detector is a single port superconducting resonator with frequency $\w_a$,  capacitively coupled to a double quantum dot (DQD) with a coupling $g$.  
The DQD is set to zero source-drain bias, with the lead chemical potentials set in between the ground and excited state.
{Absorption of a resonator photon cause transitions between the DQD states $\ket{g}$ and $\ket{e}$, resulting in current flow.}
Lower left box: Energy levels (horizontal lines) and transition processes (single directed arrows) in the resonator-DQD system.  A single photon received by the resonator ($\ket{g}\ket{n+1}$) can be emitted at the rate $\kap$ or be absorbed by the DQD at the rate $g^2/\G_2$, exciting the metastable state $\ket{e}\ket{n}$ and then decaying back to the system ground state $\ket{g}\ket{n}$ at the inelastic decay  rate $\G_1$. 
}
\label{fig:detector}
\end{center}
\end{figure}

\section{Model\label{sec:model}}

We {consider} a microwave photon detector consisting of a superconducting microwave resonator coupled to a DQD, which receives photons from  a semi-infinite  microwave transmission line, as shown in \fig{fig:detector}. 
The DQD is operated in the ``pumping" configuration, with zero source-drain bias and near the charge transition between the  charge states 
${\ket{L}=\ket{N+1,M}}$ and ${\ket{R}=\ket{N,M+1}}$, where $\ket{N,M}$ denotes $N$ ($M$) electrons in the left (right) dot.
The electron Hamiltonian governing behavior of the dot with $M+N+1$ electrons is 
\begin{equation}
H_{\rm{DQD}}=\hbar\frac{\e\tau_z+2t_c\tau_x}{2}=\frac{\hbar\w_\s\s_z}{2} , 
\end{equation}
where $\tau_{x,z}$ are the Pauli matrices in the charge basis, $\ket{L,R}$, while ${\s_z=\ket{e}\bra{e}-\ket{g}\bra{g}}$ is the diagonal Pauli matrix in the basis of eigenstates of $H_{\rm{DQD}}$:
\begin{align}
\begin{pmatrix}|g\>\\|e\>\end{pmatrix}&=
\begin{pmatrix}
\cos{\theta\over2} &  -\sin{\theta\over2}   \\
\sin{\theta\over2} &    \cos{\theta\over2}
\end{pmatrix}
\begin{pmatrix}|L\>\\|R\>\end{pmatrix}
\label{eq:ChargeStates}
\end{align}     
We introduced the following notations: $\e$ is the DQD bias from charge degeneracy point (voltage bias across the dots), $t_c$ is the interdot tunnel coupling, and ${\w_\s=\sqrt{4t_c^2+\e^2}}$ is the DQD excitation energy and ${\cos{\theta}=-\e/\omega_\sigma}$.

We consider zero bias across the leads, when 
electrons in the excited state can tunnel out incoherently to the leads,  resulting in the ``empty" state $\ket{0}=\ket{N,M}$, and then, the ground state $\ket{g}$ can be loaded by electron tunneling from the leads.  The Hamiltonian describing tunneling to the leads is given by
\[H_\G=\sum_{i=L,R}\sum_\mbk \left(\e_\mbk c^\dag_{i\mbk}c_{i\mbk}+ t_{i}c^\dag_{i\mbk}d_i+t^*_{i}c_{i\mbk}d_i^\dag~\right) ,\]
where $d_i^\dag=\ket{i}\bra{0}$, $c_{i\mbk}^\dag=\ketbra{\mbk}{i}$, $\e_k$ is the dispersion for an electron state $\ket{k}$ with wave vector $\mbk$ in the leads and $\ket{i}$ is a state in the L/R dot. The resulting incoherent tunneling rates from $\ket{e}$ to $\ket{0}$ ($\G_{0e}$) and from  $\ket{0}$ to $\ket{g}$  ($\G_{g0}$)  are given by  \cite{xuPRB13,kulkarniPRB14,gurvitzPRB96}
\begin{align}
\G_{0e}&=\G_L\cos^2(\theta/2)+\G_R\sin^2(\theta/2)\nn
\G_{g0}&=\G_L\sin^2(\theta/2)+\G_R\cos^2(\theta/2)
\label{eq:leads}
\end{align}
where $\G_{R(L)}$ are  tunneling rates to the right (left) leads.  

Taking into account opposite currents due to loading from and tunneling to the leads, the {time-averaged} electron current through the DQD can be written as
\begin{equation}
{\expval{I}}=e\G_R \left\{
\expval{p_e}|\<R|e\>|^2-\expval{p_0}|\<R|g\>|^2\right\},
\label{eq:I}
\end{equation}
where  ${p_\al=\ket{\al}\bra{\al}}$ is the projection operator to state $\ket{\al}$, $\expval{\dots}$ denotes the quantum mean value and time average. In the absence of photon excitation of the dot and at temperature ${T=0}$ {\footnote{When thermal broadening in the leads become comparable to the photon energy, there is current in the absence of photon excitation (dark counts), due to electrons that tunnel from the lead to the excited state, transition to the ground state by emiting a photon, and then tunneling from the ground state to the lead.  However, this effect can be made negligible by lowering the temperature to $40$ mK= $0.8$ GHz $h/k_B$, an order of magnitude lower than the photon energy we consider.}},
the DQD is in the ground state, ${\expval{p_g}=1}$ and ${\expval{p_e}=\expval{p_0}=0}$, and no current flows. However, photons in the microwave resonator cause transitions between the ground and excited states of the DQD, resulting in a finite current.  Photon arrivals can thus be detected by measuring the DQD source-drain current.

  

The Hamiltonian of non-interacting photons and the DQD in the rotating frame at the frequency of input photons $\w_{\rm in}$ are
\begin{align}
H_{\rm free}&=\int_{-\infty}^\infty {d\w\over2\pi}\hbar\w b^\dag_\w b_\w+\hbar\Delta_{ab}\left(a^\dag a+\onehalf\right),\label{eq:Hfree}\\
\tilde H_{\rm DQD} & = \frac{\hbar\D_{\s b}}{2}\s_z.
\label{eq:H2}
\end{align}
where ${\Delta_{ab}=\w_a-\w_{\rm in}}$  (${\D_{\s b}=\w_\s-\w_{\rm in}}$) is the detuning of the resonator frequency (DQD) $\w_a$ ($\w_\s$) from $\w_{\rm in}$.   Here, we used the mode expansions
\begin{align}
V_{\rm TL}&=\int_{-\infty}^\infty{d\w\over2\pi}\sqrt{\hbar\w Z_{\rm TL}\over2}(b_\w+b_\w^\dag)\label{eq:VTL}~.\\
V_a&= \sqrt{\hbar \w_a \over2 C_a}(a{+}a^\dag)~,\label{eq:Va}
\end{align}
for the resonator ($V_a$) and transmission line ($V_{\rm TL}$) voltage operator at the coupling capacitor $C_\kappa$, where $a$ is the resonator photon annihilation operator  and {$b_\w$} is the transmission line  photon annihilation operator  with frequency $\w$ 
\footnote{The resonator mode $a$ is dimensionless, while the  transmission line modes have units of $b_\omega\sim \sqrt{\rm time}$ and $b_{\rm in }(t)\sim 1/\sqrt{\rm  time}$.}.
Furthermore,  ${Z_{\rm TL}=\sqrt{l/c}}$ is the characteristic impedance of the line,  $l$ is the inductance per unit length, $c$ is the capacitance to ground per unit length, and $C_a$ is resonator capacitance.  We assume the transmission line dispersion  ${\w_k=vk}$, where $v=1/\sqrt{lc}$ is the group velocity.

The  resonator-DQD coupling is described by the Jaynes-Cummings Hamiltonian: 
\begin{equation}
H_{\rm{JC}} =  \hbar g (\sigma^\dagger a+a^\dagger \sigma),
\end{equation} 
where  ${\s=\ket{g}\<e|}$ is the DQD lowering operator. Here, the interaction strength ${g=g_0\sin\theta}$ is defined by the dipole matrix element in the basis of energy eigenstates, and
\[g_0=\omega _a \beta _g \sqrt{\frac{\pi  Z_a}{R_Q}}~,\]
where $Z_a$ is the resonator characteristic impedance,  $R_Q=h/2e^2=12.9$ k$\W$ is the resistance quantum, $\beta_g=C_g/(C_g+C_\Sigma)$, $C_g$ is the gate capacitance between the resonator and the DQD, and $C_\Sigma$ is the total capacitance of the DQD \cite{childressPRA04}.

The photon mode in the resonator is driven by photons exchange with the transmission line.  Assuming a weak, local  coupling capacitance $C_\kap$ between transmission line photons and the resonator, the interaction Hamiltonian in the rotating wave approximation is given by
\begin{subequations}
\begin{align}
H_{\rm TL}&=C_\kap V_aV_{TL}=\hbar\sqrt{\kap} \int {d\w\over2\pi}\qty[b^\dag_\w a+a^\dag b_\w],\label{eq:HTL}
\end{align}
where 
\begin{equation}
\kappa=C_\kappa^2 \w_{\rm in}^3 \frac{Z_{\rm TL}Z_a}{4}~,
\end{equation}
\end{subequations}
{is the photon leakage rate from the resonator, $Z_a=\sqrt{L_a/C_a}$ is the impedance of the coplanar waveguide resonator, and $L_a$ is the resonator inductance.  We notice that the resonator quality factor can be expressed as
\[Q=\frac{\omega _{\text{in}}}{\kappa }=\frac{4 Z_{\text{TL}}}{Z_a \left(\omega _{\text{in}} \tau_{\text{RC}}\right)^2}\]
 where $\tau_{\rm RC}=C_\kap Z_{\rm TL}$ is the recharging time of the coupling capacitor $C_\kap$.}
To finalize the description of the model, we present the full Hamiltonian of the system: 
\begin{subequations}
\begin{eqnarray}
H & =H_{sys}+H_{\rm{TL}}+H_\G+H_\g \, \label{eq:Htot}\\
H_{sys} & =\til{H}_{DQD} +H_{JC}+H_{free}~,
\end{eqnarray}
\end{subequations}
In \eq{eq:Htot}, $H_\g$ is the Hamiltonian describing the DQD dissipative environment, which consist of voltage fluctuations and phonons.

\section{equations of motion\label{sec:EOM}}

We will employ input-output theory \cite{gardinerBook04,collettPRA84,wallsBook2012,clerkRMP10} to model the resonator-transmission line interaction.  {This formalism will enable us to optimize the quantum efficiency of photon detection including interference effects between the  microwave signals reflected by the resonator-DQD system, which is not captured by density matrix master equations}.  The key assumptions in this formalism is the rotating wave   and Markov approximation.  
 
The equation of motion for the transmission line modes that follows from the Hamiltonians in \eq{eq:HTL} and \eq{eq:Hfree} can be solved analtyically to yield
\ben
b_\w(t)=e^{-i\w(t-t_0)}b_\w(t_0)+{i} \sqrt{\kap}\int_{t_0}^t\,dt'e^{-i\w(t-t')}a(t')
\label{eq:b}
\een
which is a solution that can be specified by initial $t_0=t_i<t$ or final condition $t_0=t_f>t$,  at time $t_i$ ($t_f$) long before (after)  the transmission line photons interact with the resonator.  The input field in the rotating frame is defined by the initial field configuration
as 
\begin{align}
b_{\rm in}(t)&=
\int_{-\infty}^\infty\frac{d\w}{2\pi}b_\w(t_i)e^{-i\w (t-t_i)}~.
\label{eq:bin}
\end{align}
We will consider the case where the input field is formed by a continuous flux of photons in a narrow spectral band around $\w_{\rm in}$, {$b_\w(t_i)\propto\de(\w-\w_{\rm in})$.} 
The flux of incoming photons is given by ${\dot{N}=\langle b_{\rm in}^\dag(t) b_{\rm in}(t)}\rangle$.
In terms of experimental parameters, $\Ndot=vn_b/L=n_b/Z_{\rm TL}cL$, where $n_b$ is total input photon number of the waveguide and the transmission line impedance is $Z_{\rm TL}=50 \W$.

Using the solution \eq{eq:b} specified by $b_{\rm in}$ leads to the Heisenberg-Langevin equations for the system operators  $\{X\}=\{a,\s,\s_z,p_0\}$ given by  \cite{clerkRMP10,wallsBook2012,gardinerBook04}
\begin{align} 
\dot{X}&={i\over\hbar}[H_{sys},X]+(D_\g+D_\G) X\nn
&+(\kap/2{-i}\sqrt{\kap}b^\dag_{\rm in})[X,a]+[a^\dag,X](\kap/2+{i}\sqrt{\kap}b_{\rm in})
\label{eq:HL}
\end{align} 
where 
 $\mcal{D}(A)$ is the Linblad superoperator defined by $\mcal{D}(A)\rho=A\rho A^\dag-(A^\dag A\rho+\rho A^\dag A)/2$. Here, $D_\G$ is the dissipative operator describing  incoherent tunneling to the leads, given by
\begin{align}
D_\G&=\G_{0e}\mcal{D}(\ket{0}\<e|)+\G_{g0}\mcal{D}(\ket{g}\<0|)
\label{eq:D1}
\end{align}
and $D_\g$ describes relaxation of the DQD eigenstates
\begin{align}
D_\g&=\g_\up(\e) \mcal{D}(\s^\dag )+\g_\down(\e)\mcal{D}(\s)+\frac{\g_\phi(\e)}{2} \mcal{D}(\s_z)\,.
\label{eq:Dg}
\end{align}
In \eq{eq:Dg},  $\g_\up(\e)$, $\g_\down(\e)$ and $\g_\phi(\e)$ are bias dependent charge excitation, relaxation, and pure dephasing rates, respectively.  Analytic expressions for these rates and a plot of their bias dependence are given in Appendix \ref{sec:noise}.  They are caused by voltage fluctuations, which simply adds a classical noise component to the bias $\e\to\e+\de\e(t)$, where $\de\e(t)$ has a $1/f$ noise spectrum [cf.~\eq{eq:epsnoise}], and phonons, which  couple to the DQD charge density in a manner similar to the coupling to the resonator field, [cf.~ \eq{eq:Hph}] \cite{kulkarniPRB14}.  


From \eq{eq:HL} and \eq{eq:D1}, the equations of motion for the photon and DQD operators are \cite{gardinerBook04}
\begin{subequations} \begin{align} 
\dot{a}&=-\left(\frac{\kap}{2}+i\Delta_{ab}\right)a-ig\s{+i}\sqrt{\kap}b_{\rm in}\label{eq:a}\\
\dot{\s}&=-({\G_2}+i\D_{\s b})\s{+}iga\s_z\label{eq:s}
\end{align} \end{subequations} 
where
\ben\G_2=\g_\phi+\frac{\g_1+\G_{0e}}{2}\label{eq:gamma2}\een
is the transverse relaxation rate. In addition to the usual contribution from charge qubit contributions from the pure dephasing rate $\g_\phi$ and inelastic decay from $\ket{e}\to\ket{g}$ at the rate ${\g_1=\g_\up+\g_\down}$, there is lifetime broadening due to incoherent tunneling to the leads  at the rate $\G_{0e}$. 

The equations of motion the DQD polarization and ``empty" state  projection operators are  
\begin{subequations} \begin{align} 
\dot{\s}_z&=2ig(a^\dag\s-\s^\dag a)\label{eq:sz}\\
&+\left(\gamma _--\frac{{ \Gamma_{0e}}}{2}\right) p_{\sigma }-\left(\gamma _1+\frac{{ \Gamma_{0e}}}{2}\right) \sigma _z-p_0 \Gamma _{{g0}}\,,\nn
\dot{p}_0&=\G_{0e}\frac{p_\s+\sigma _z}{2}-\G_{g0}p_0\label{eq:p0}
\end{align} \end{subequations} 
where $p_\s=\ket{e}\<e|+\ket{g}\<g|$ is the projection operator into the charge qubit subspace, determined by the constraint the  $p_0+p_\s=\ket{e}\<e|+\ket{g}\<g|+\ket{0}\<0|=1$, which represent conservation of probability in the DQD state space.

\section{Steady-state solution\label{sec:SSS}}

We will compute the detection efficiency for the case of continuous flux of photons \cite{hadfieldNATP09} using the steady state solution to the equations of motion. Here, we present the steady state solution for the polarization $m_z\equiv \expval{\s_z}$, for a purely quantum input field with zero classical component $\expval{b_{\rm in}}=0$, and then relate $m_z$ to the detection efficiency of the device \footnote{When $\expval{b_{\rm in}}\neq 0$, we have to keep the term $\expval{a}\s_z$ in the equation for $\s$.}.

We eliminate  the ``empty" dot operator $p_0$ using the steady state solution of Eq.~\eqref{eq:p0}, which yields
\ben
p_\s=\frac{2 \Gamma _{{g0}}-\Gamma _{0 e} \sigma _z}{\Gamma _{0 e}+2 \Gamma _{{g0}}}\,.
\label{eq:psigss}
\een
Then, substituting \eq{eq:psigss} to \eq{eq:sz} yields an effective equations of motion for the polarization
\begin{subequations}
\begin{align}
\dot{m}_z&=\G_1( m_z^0-m_z)+2ig\expval{a^\dag\s-\s^\dag a}\,,\label{eq:sz1}\\
\Gamma_1& =\g_1+\frac{2 \G_{0e} \Gamma _{g0}-\gamma _s \G_{0e}}{\G_{0e}+2 \Gamma _{g0}}\label{eq:gamma1}\\
 m_z^0&=-\frac{(\gamma_s+\G_{0e}) \Gamma _{{g0}}}{(\gamma _1+\G_{0e}) \Gamma _{{g0}}+\G_{0e}(\gamma _1-\g_s)/2}\label{eq:mz0}
\end{align}
\end{subequations}
where $\G_1$ is the effective depolarization rate and ${\g_s=\g_\down-\g_\up}$ is the phonon-induced spontaneous emission rate.  

At this point, the three level DQD system consisting of $\{\ket{e},\ket{g},\ket{0}\}$ is reduced to a two level system described by Eqs.~\eqref{eq:s} and \eqref{eq:sz1} with longitudinal and transverse relaxation rates $\G_1$ and $\G_2$, respectively.  The equilibrium polarization $m_z^0$  gives the value of $m_z$ in the absence of coupling to resonator photons, $g=0$. The effective DQD excitation, relaxation, and dephasing rates are $\G_\up=\G_1(1+ m_z^0)/2$ and $\G_\down=\G_1(1- m_z^0)/2$, and $\G_\phi=\G_2-\G_1/2$, respectively.  

The steady state photon and DQD operators satisfy
\begin{subequations}\begin{align} 
a&=\chi_a({\sqrt{\kap}b_{\rm in}-g\s})\label{a2}\\
\sigma&=g\chi_\s (a  \sigma _z)\label{eq:sm2}
\end{align} \end{subequations}
where $\chi_a=(\Delta_{ab}-i\kap/2)^{-1}$ and $\chi_\s =(\D_{\s b}-i\Gamma_2)^{-1}$ are the resonator and DQD susceptibilities \cite{clerkRMP10}. To solve these equations, we apply the mean field approximation to \eq{eq:sm2} by taking $a \s_z\to am_z$, and substitute \eq{eq:sm2} into \eq{a2} to find the resonator photon field
\begin{subequations}\begin{align} 
a&=\chi_{ab}b_{\rm in}\,,\\
\chi_{ab}&=\frac{\sqrt{\kap}}{\Delta_{ab}-i\kappa /2+g^2m_z\chi_\s}\, .\label{eq:a2}
\end{align} \end{subequations}
This solution, together with the steady state solution of \eq{eq:sz1} yields the mean field equation 
\begin{subequations}\begin{align} 
m_z&=\frac{ m_z^0}{1+G\expval{n_a}}\,, \label{eq:meanMz} \\
\expval{n_a}&=|\chi_{ab}|^2\expval{b^\dagger_{\rm in}b_{\rm in}},\label{eq:na} 
\end{align}\end{subequations}
where  ${G=(4g^2/\G_1){\rm Im}\chi_\s}$ and $n_a=a^\dag a$ is the resonator photon number operator.
  
Due to the dependence on $m_z$ in  \eq{eq:a2}, the mean field equation~\eqref{eq:meanMz} is a cubic equation for $m_z$ \cite{wallsBook2012}.  
However, a simple estimate based on the perturbative parameter $G\leq g^2/\G_1\G_2\simeq 10^{-4}$ will show that it is sufficient to take ${m_z=m_z^0}$ in \eq{eq:a2}, so that \eq{eq:meanMz}  yields explicit solution for $m_z$.
The photon-induced polarization is
\ben \D m_z=m_z-m_z^0=-m_z^0 \frac{G\expval{n_a}}{1+G\expval{n_a}}~,\label{eq:deltaMz}\een
which, to leading order yields $\D m_z\approx|m_z^0|G\expval{n_a}\simeq G\dot N/\kap$, where we kept  $\expval{n_a}\simeq\dot N/\kap$  to $\order{1}$.  Substituting this in \eq{eq:a2} yields the leading order correction to the photon number $\expval{\D n_a}\simeq C G (\dot N/\kap)^2$, where \ben {C=g^2/\G_2\kap}\label{eq:C}\een
is the cooperativity.  As shown in section \ref{sec:eff}, $C=\order{1}$, and, since we will consider input flux $\Ndot$ and leakage rate $\kap$ in the MHz range, $\expval{\D n_a}=\order{G}$, which results in subleading term of $\order{G^2}$ in $\D m_z$.
The $\order{G}$ estimate for  the induced polarization will be verified numerically below (see \fig{fig:secV}d).  The linearization of  \eq{eq:sm2} is justified by the same perturbative expansion.

\section{Quantum efficiency\label{sec:eff}} 

The detector efficiency is defined by the ratio of the steady state mean DQD current  per input photon flux,
\ben
\eta=\frac{|\expval{\D I}|}{e\Ndot}
\label{eq:eta}
\een
where  $e$ is the electron charge, $\expval{\D I}=\expval{I-I_0}$ is the mean polarization and current  caused by absorbing photons and  $\expval{I_0}$ are the dark counts due to the current at zero {photon} input flux. 
The electron current \eq{eq:I} can be expressed in terms of the polarization as
\ben
{\expval{I}}=-\frac{e\G_R(1+m_z)}{\G_{0e}+2 \Gamma _{{g0}}}\qty( \G_{0e}\cos^2\frac{\theta}{2}-\Gamma _{g0}
\sin^2\frac{\theta}{2})\,.
\label{eq:barI1}
\een
As one would expect, the current is proportional the probability for the DQD to be excited, $\expval{p_e}=(1+m_z)/2$. The last  factor in Eq.~\eqref{eq:barI1} takes into account the cancellation of the left and right moving {electrons}.

We will henceforth consider symmetric dot-lead tunneling at the rate $\G_l=\G_L=\G_R$. Then $\G_{0e}=\G_{g0}=\G_l$,  the photon induced current is
\ben
{\expval{\D I}}=-e{\G_l\cos \theta\over3}\D m_z~,
\label{eq:deltaI}
\een
and Eqs.~\eqref{eq:mz0} and  \eqref{eq:gamma1} become
\begin{align}
\Gamma_1&=\g_1+\frac{2 \Gamma _l-\g_s}{3}\,,\nn 
m_z^0&=-\frac{\gamma_s+\Gamma_l }{3\g_1/2-\g_s/2+\Gamma_l}\,.
\label{eq:mz0a}
\end{align}

We will be interested in the linear regime with respect to $\Ndot$, where single photon detection occurs.  This regime coincides with the leading order expansion of $G\expval{n_a}$ in \eq{eq:deltaMz}, where the efficiency is given by
\ben
\eta={\frac{4}{3}\frac{\kap g^2(\G_l/\G_1)\abs{\cos \theta {\rm Im}\chi_\s }}{(\Delta_{ab}-g^2 |m_z^0| {\rm Re}\chi_\s )^2+(\kappa/2+g^2|m_z^0|{\rm Im}\chi_\s)^2}}\,.\label{eq:eta1}
\een

\begin{figure}[t]
\begin{center}
\includegraphics[width=.8\linewidth]{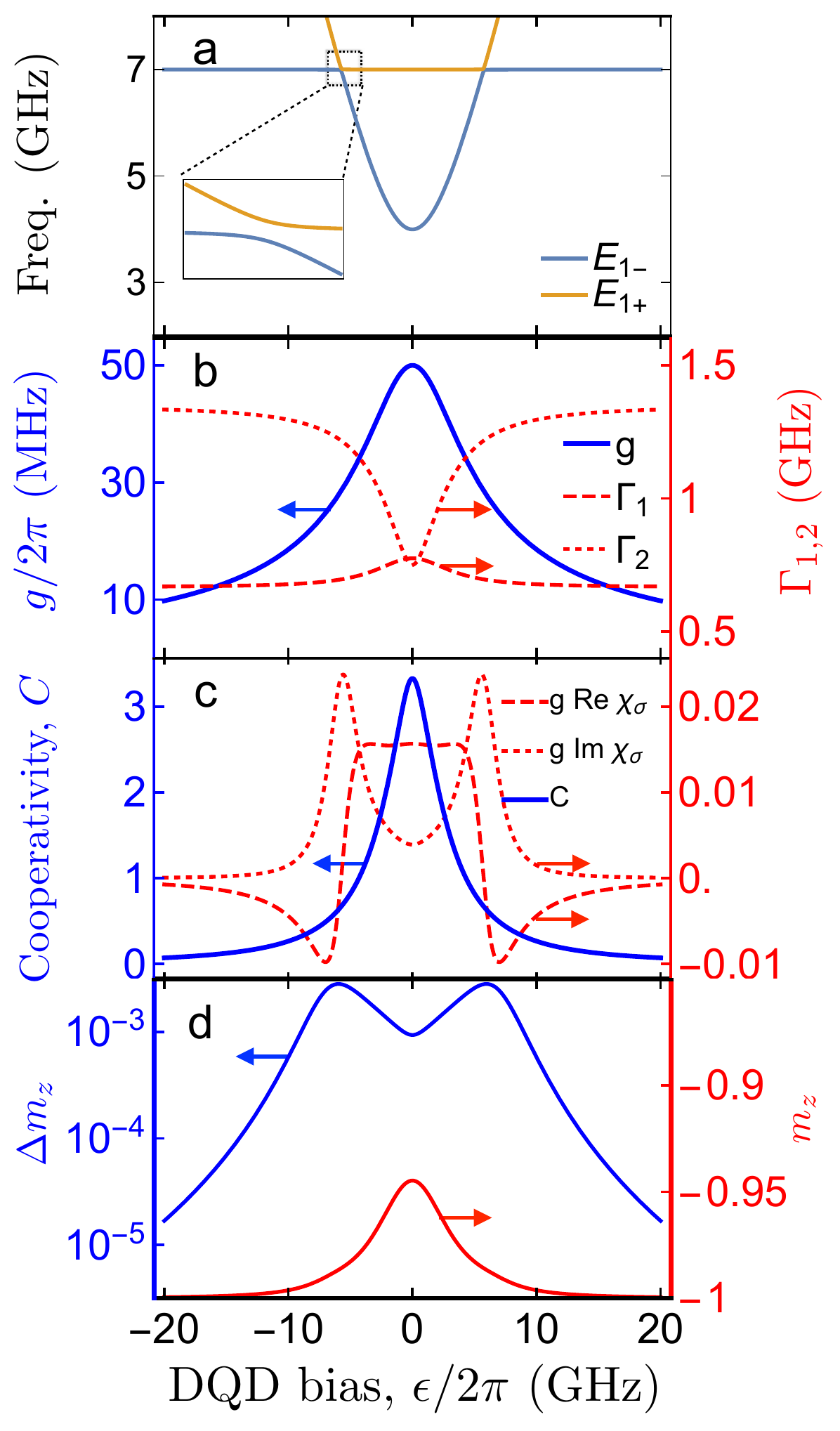}
\caption{As a function of the double dot bias $\e$, for  ${(t_c,g)/2\pi =(2, 0.05)}$ GHz and ${\Ndot=1}$ MHz:
(a) Jaynes-Cummings energy levels $E_{1\pm}$ of the DQD-resonator system.
(b) DQD-resonator coupling ${g=g_0\sin\theta}$ for ${g_0/2\pi=50}$ MHz, transverse relaxation rate $\G_2$ and depolarization rate $\G_1$ of the effective two level DQD.
(c) Cooperativity of the DQD-resonator system ${C=g^2/\G_2\kap}$; the real and imaginary parts of the DQD susceptibility $\chi_\s$.
(d) DQD polarization $m_z$ and photon-induced polarization $\D m_z$.}
\label{fig:secV}
\end{center}
\end{figure}

Next, we analyze quantities that characterize the DQD-resonator coupling, dissipation, and response as a function of bias, see \fig{fig:secV}a-c. 
As a point of reference,  we show in \fig{fig:secV}a the Jaynes-Cummings energy levels in the zero and one photon subspace.
\[{E_{1,\pm}=\hbar\w_a\pm{\hbar\over2}\sqrt{4g^2+\D_{a\s}^2}}~,\]
where ${g=g_0\sin\theta}$ and we take the parameters ${\w_a/2\pi= 7}$ GHz and ${g_0/2\pi=50}$ MHz, which is a value that has been reported in experiments \cite{freyPRL12,dengPRL15,viennotPRB14}.  
In \fig{fig:secV}, we choose a tunnel coupling of ${t_c/2\pi=2}$ GHz, but we will optimize this parameter below.
{Note that, since $\G_2\gg g$, the resonances at the Jaynes-Cummings energies are destroyed;} the subspace of the resonator-DQD system with $n$ photons is specificed by the uncoupled basis  $\{\ket{g}\ket{n},\ket{e}\ket{n-1}\}$, where $\ket{e}$ denotes the DQD excited state that is strongly broadened  with a linewidth $\G_2$, as shown in \fig{fig:detector} (lower left box).

We use DQD relaxation parameters appropriate for silicon DQD \cite{wangPRL13,peterssonPRL10}: phonon noise spectral density ${\g_{\rm ph}/2\pi=0.1}$ GHz, quasistatic bias noise variance ${\e_{\rm rms}/2\pi=1.2}$ GHz, and take dot-lead tunnel rate ${\G_l/2\pi=1}$ GHz. 
The DQD-resonator  coupling $g$ (\fig{fig:secV}b) has a broad peak centered at $\e=0$, due to a strong dipole moment at the charge degeneracy point,
while the transverse relaxation rate $\G_2$ has a minimum due to a sweet spot where dephasing is to the first order insensitive to quasistatic bias noise.  
These effects combined lead to a broad peak around ${\e=0}$ in the cooperativity ${C=g^2/\G_2\kap}$, shown in \fig{fig:secV}c, indicating strong DQD-resonator interaction.
The strong charge dipole also increases transitions driven by phonons and $1/f$ charge noise, leading to a peak in the equilibrium polarization of ${m_z^0\simeq-0.95}$ at ${\e=0}$ (\fig{fig:secV}d) and a peak in the depolarization rate $\G_1$ (\fig{fig:secV}b), which is otherwise  dominated by dot-lead  tunneling.   The real and imaginary part of $\chi_\s$, which modify the effective resonator frequency and decay rate, respectively, are plotted in \fig{fig:secV}c

\begin{figure}[t]
\begin{center}
\includegraphics[width=\linewidth]{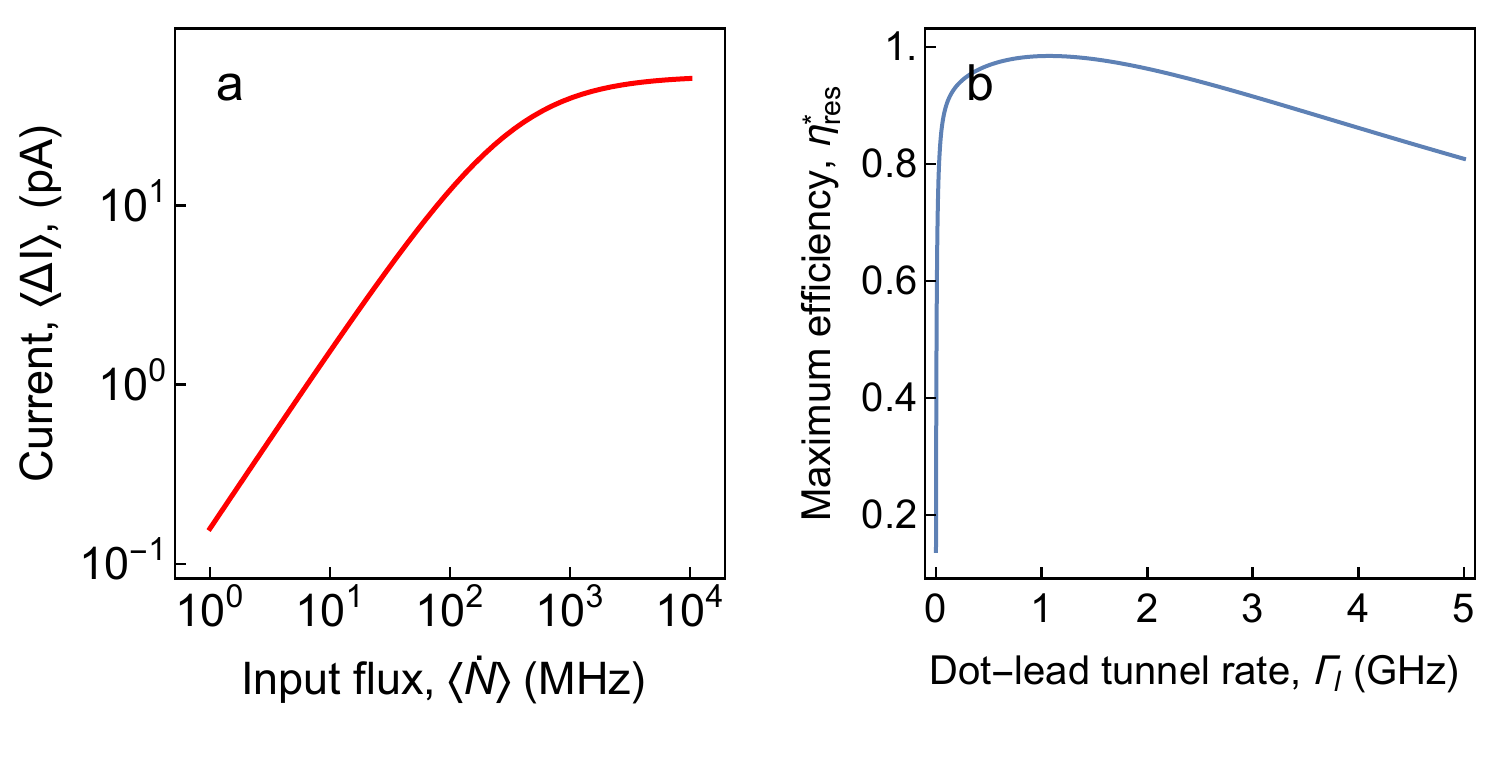}
\caption{{For optimal parameters ${\kap/2\pi=76}$ kHz and ${t_c/2\pi=0.5}$ GHz, on resonance $\D_{ab}=0$: (a) Logarithmic plot of the photon-induced current $\expval{\D I}$ as a function of input flux $\Ndot$ for $\G_l=1$ GHz. 
(b) Maximum efficiency as a function of the incoherent dot-lead tunnel rate $\G_l$ with $\Ndot=1$ MHz.}}
\label{fig:sat}
\end{center}
\end{figure}

We conclude this section by considering the detector response as a function of input flux $\Ndot$ and dot-lead tunnel rate $\G_l$.
Nonlinear response at large flux will cause $\D I$ to saturate and sets the detector dead time \cite{hadfieldNATP09}. 
At the same time, a sufficiently large flux is necessary for the current ${\D I=\eta e\Ndot}$ to be measurable.
Figure~\ref{fig:sat}a shows the photon-induced current 
\[\D I=0.16\eta \Ndot~{\rm pA/MHz}\]
as a function of $\Ndot$ at $\G_l=1$ GHz using \eq{eq:etamax} for the optimal efficiency $\eta^*_{\rm res}$ found below. The response is linear up to ${\Ndot=100}$ MHz, and saturates as the flux approaches the effective inelastic decay rate ${\Ndot\simeq\G_1}$.  
\fig{fig:sat}b shows  $\eta^*_{\rm res}$ as a function of the lead tunnel rate $\G_l$ at fixed $\Ndot=1$ MHz,
The optimal rate occurs near $\G_l=1$ GHz, and is determined by the competition between two effects: when $\G_l$  is too low, the electron relaxes back to $\ket{g}$, but when $\G_l$  is too high, the efficiency suffers due to level broadening, $\G_2\propto\G_l$, see \eq{eq:gamma2}. 

Maximum photon absorption by the DQD occurs on resonance, as shown by the peaks at $\e_{\rm res}$ in the photon-induced DQD polarization $\D m_z$ plotted  in \fig{fig:secV}d, which are accompanied by minima in photon number with ${n_a\simlt1}$, indicating perfect photon to electron conversion.
Note that induced polarization is very small, ${\D m_z(\e_{\rm res})\simeq10^{-4}\ll m_z^0}$,  which justify our approximation of taking ${m_z=m^0_z}$ in \eq{eq:a2}, and agrees with our previous estimate below \eq{eq:C} that $\D m_z=\order{G}$.

Figure \ref{fig:eff}a shows the photon detector efficiency as a function of bias for $\kap=1$ MHz, computed using \eq{eq:deltaMz}, \eqref{eq:eta} , and \eqref{eq:deltaI}.     When the charge transition is sharp, at  ${t_c/2\pi=0.5}$ GHz, a double-peak behavior emerges:
one peak is due to resonance and the other is due to the competition between enhanced cooperativity and cancellation of left and right moving currents, which goes as $\sin^2\theta\cos\theta$. As expected, the maximum efficiency occurs on resonance:  ${\eta_{\rm res}\simeq80\%}$ at $\e=\e_{\rm res}$ and ${t_c/2\pi=1.5}$ GHz. This efficiency will be further optimzied in section \ref{sec:optimal}.

\section{Reflected Signal\label{sec:ref}}

\begin{figure}[t]
\begin{center}
\includegraphics[width=0.75\linewidth]{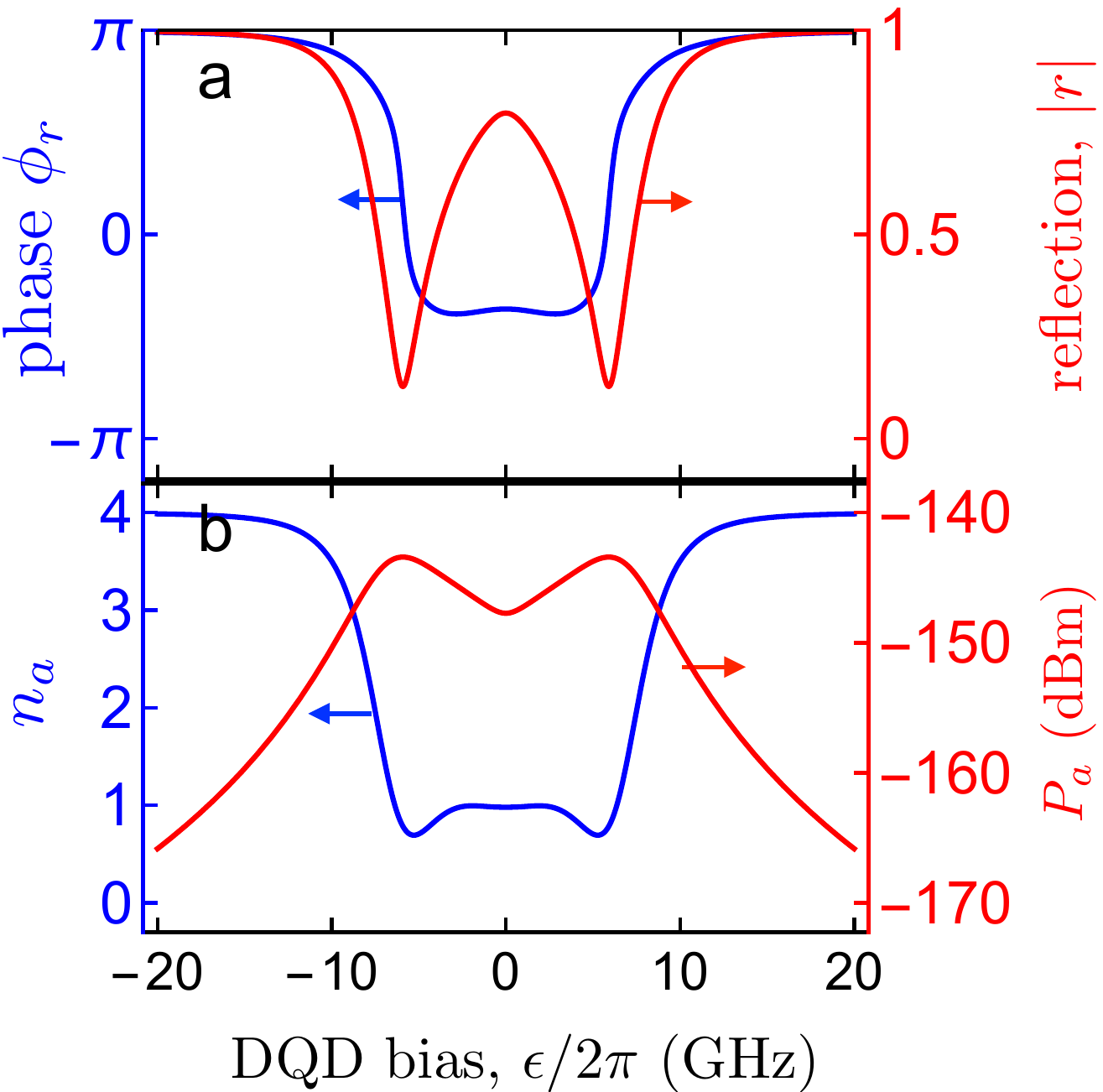}
\caption{As a function of the double dot bias $\e$, for  ${(t_c,g)/2\pi =(2, 0.05)}$ GHz and ${\Ndot=1}$ MHz:
(a) {Magnitude $|r|$ and phase $\phi_r$ of the reflection coefficient.}
(b) Photon number $n_a$ and power absorbed $P_a$.}
\label{fig:secVI}
\end{center}
\end{figure}

The field of the transmission line, see Eq.~\eqref{eq:b}, can be described in terms of its final configuration at $t_0=t_f>t$.  In this case, we can introduce the output field as a counterpart of the input field, Eq.~\eqref{eq:bin}:
\begin{equation}
b_{\rm out}(t)=\int_{-\infty}^\infty\frac{d\w}{2\pi}b_\w(t_f)e^{-i\w (t-t_f)}~.
\end{equation}
The equation of motion for the resonator field has a structure similar to that of 
Eq.~\eqref{eq:a}, but an opposite sign in front of the term describing spontaneous emission:
\begin{equation}
\dot{a}=\left(\frac{\kap}{2}-i\Delta_{ab}\right)a-ig\s{+i}\sqrt{\kap}b_{\rm out}
\end{equation}
By subtracting the above equation from Eq.~\eqref{eq:a}, 
we obtain the relation between input and output modes \cite{clerkRMP10},
 \ben
{b_{\rm out}(t)=b_{\rm in}(t)+{i}\sqrt{\kap}a(t)}~,
\label{eq:bout}
\een 
where on the right hand side, the first term is the reflection of the input field and the second is the field radiated by the resonator.  The reflection coefficient $r$ is defined by ${b_{\rm out}=rb_{\rm in}}$. 

A system of coupled DQD and microwave resonator can be also used to control the output microwave field \cite{schiroPRB14,goldsteinPRL13}.  Here, we briefly analyze the suppression of the reflected signal from the resonator when the DQD device acts as an adjustable dissipating element.
We consider the reflecting signal of microwave photons 
for input photon frequency {equal to} the resonator frequency, ${\D_{ab}=0}$, and for input flux ${\Ndot=1}$ MHz, which is well within the linear regime, as shown below. 
In \fig{fig:secVI}a, we plot the magnitude $|r|$ and phase $\phi_r$ of  the reflection coefficient, computed by using the general relation between input and output modes \eq{eq:bout}  with $a$ given by Eq.~\eqref{eq:a2}, 
which yields
\ben{r=|r|e^{i\phi_r}}=1+\sqrt{\kap}\chi_{ab}~.\label{eq:r}\een
The mean photon number ${\expval{n_a}}$ Eq.~\eqref{eq:na} and absorbed input power ${P_{a}=\hbar \omega_{\rm in}(1-|r|^2)\Ndot}$ is plotted in \fig{fig:secVI}b. {When the input frequency is on resonance with the DQD excitation at ${\e=\e_{\rm res}=\sqrt{\w^2_{\rm in}-4t_c^2}}$, reflection $|r|$ is minimal and power absorption $P_a$ is maximal.}

\section{Optimal conditions\label{sec:optimal}}

\begin{figure}[t]
\begin{center}
\includegraphics[width=0.85\linewidth]{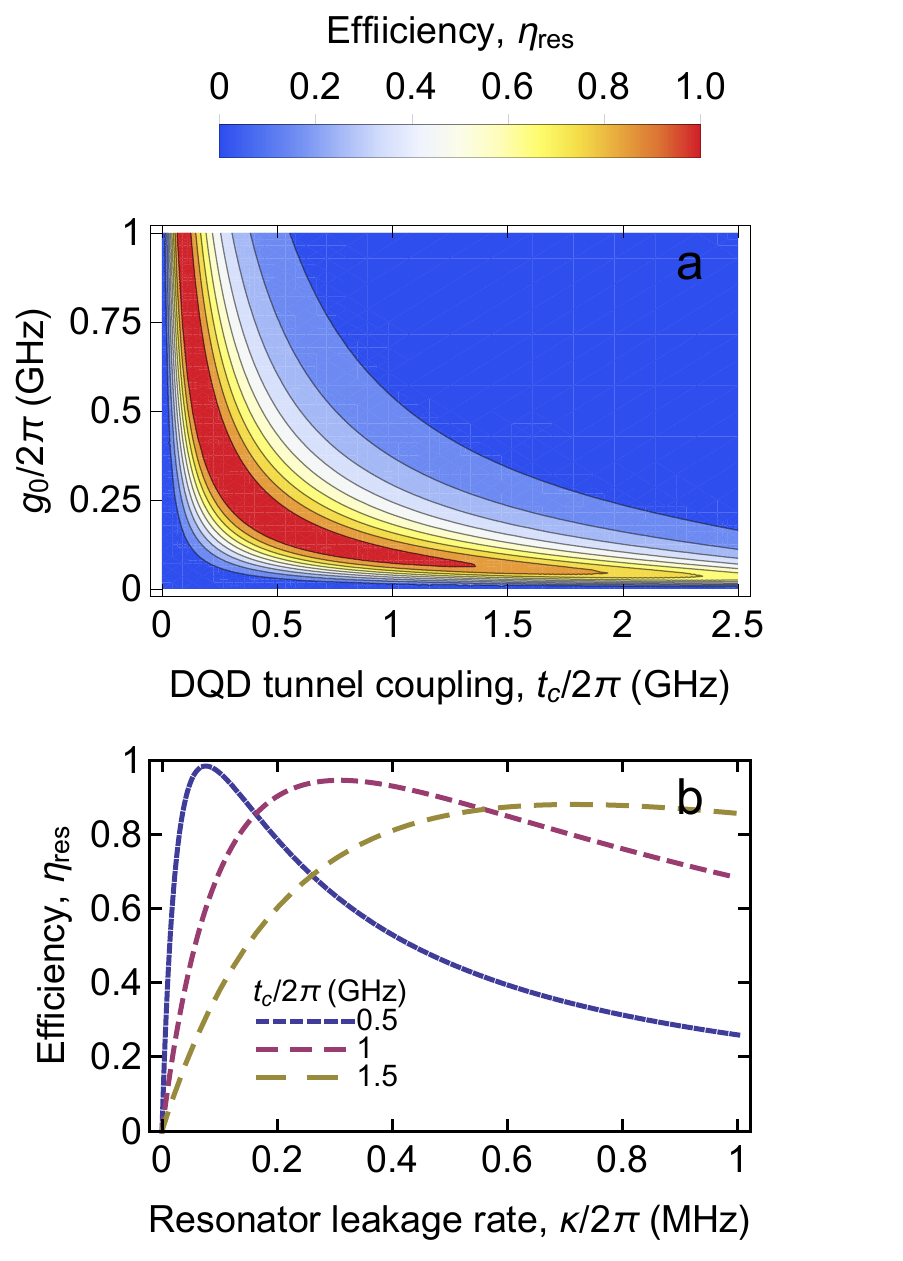}
\caption{(a) Photon detector efficiency on resonance $\eta_{\rm res}$ as a function of DQD-resonator coupling $g_0$ and interdot DQD tunnel coupling $t_c$, for ${\kap/2\pi=1}$ MHz. 
(b) Efficiency $\eta_{\rm res}$ as a function of resonator leakage rate $\kap$.}
\label{fig:optimal}
\end{center}
\end{figure}

The photon detector efficiency can be further improved by reducing the reflection of input photons, which for the parameters chosen so far is nonzero even on resonance, as shown in \fig{fig:secVI}a. 
Using \eq{eq:a2} and \eqref{eq:r},  we find $r(\e_{\rm res})=0$ when the resonator leakage rate $\kap$ matches the DQD-mediated photon dissipation rate, 
\ben
\kap=\frac{2(g_0\sin\theta_{\rm res})^2|m^0_z|}{\G_2(\e_{\rm res})}\,.
\label{eq:opt}
\een
The latter can be understood from Fermi golden rule as a transition from the one-resonator photon state $\ket{g}\ket{1}$ to the broadened DQD excited state $\ket{e}\ket{\emptyset}$ where $\ket{\emptyset}$ denotes the empty resonator state. The factor $m_z^0$ takes into account that this transition can occur only when {$\ket{g}$ ($\ket{e}$) is (un)occupied.} These transitions are illustrated in \fig{fig:detector}.  We note that \eq{eq:opt} can be expressed as the condition on the cooperativity [cf.~\eq{eq:C}]
\[{C(\e_{\rm res})={1\over2|m_z^0(\e_{\rm res})|}}\approx\onehalf~,\]
 so the optimal point does not require strong resonator-DQD coupling.

\begin{figure}[t]
\begin{center}
\includegraphics[width=0.75\linewidth]{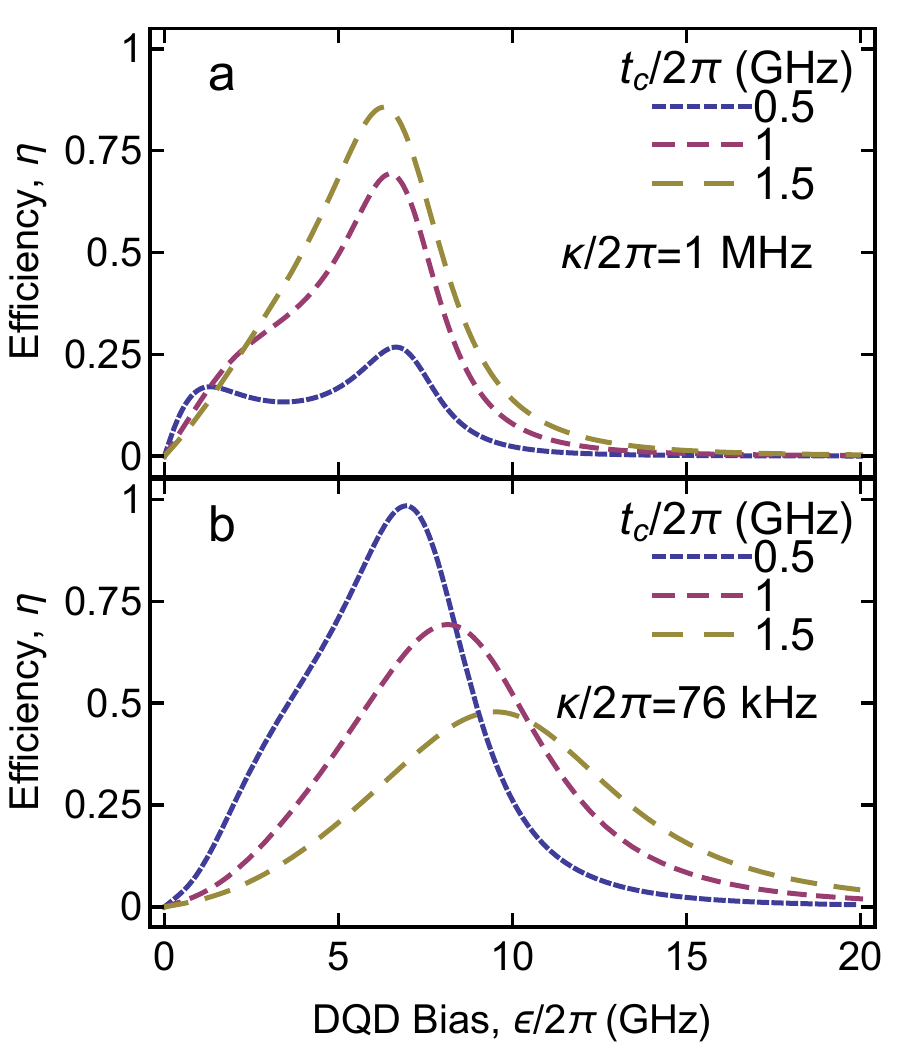}
\caption{Efficiency of photon detector  $\eta$ as a function of DQD bias $\e$, with the tunnel couplings $t_c/2\pi=(0.5,1,1.5)$ GHz and DQD-resonator coupling ${g_0=50}$ MHz, for
(a) Nominal resonator leakage rate, ${\kap/2\pi=1}$ MHz, and 
(b) Optimal resonator leakage rate, $\kap/2\pi=76$ kHz, \eq{eq:opt}.}
\label{fig:eff}
\end{center}
\end{figure}

When \eq{eq:opt} is satisfied, the efficiency on resonance is given by
\ben
\eta_{\rm res}^*=\abs{\cos\theta_{\rm res}}\frac{2\Gamma_l/3\Gamma_1}{1+2\Ndot/|m_z^0|\Gamma _1}\,.
\label{eq:etamax}
\een
For a sufficiently fast dot-lead tunnel rate ${\G_l\gg\g_{\up,\down}}$, ${\G_1\simeq3\G_l/2}$, see \eq{eq:mz0a}, so that $\eta_{\rm res}$ is limited mainly by the factor 
\ben\abs{\cos\theta_{\rm res}}=\sqrt{1-4t_c^2/\w_{\rm in}^2}~.\label{eq:cos}\een
Higher maximum efficiency $\eta_{\rm res}$ can be thus achieved by lowering $t_c$, but at the cost of increasing the optimal coupling $g^*_0$, due to the reduction in the DQD-resonator coupling by ${\sin\theta_{\rm res}=2t_c/\w_{\rm in}}$. 
This behavior is shown in \fig{fig:optimal}a, where we plot efficiency on resonance  $\eta_{\rm res}$  as a function of $t_c$ and $g_0$.

The optimal regime defined by \eq{eq:opt} can be reached by tuning DQD-resonator parameters $t_c$, $g_0$, and $\kap$.  At  $\kappa=1$ MHz, it can satisfied at ${(t_c,g^*_0)\simeq2\pi(0.5,0.2)}$ GHz, but this $g^*_0$ is larger than presently available DQD-resonator couplings.  We therefore propose lowering the resonator leakage rate $\kappa$ at fixed $g_0$.   \fig{fig:optimal}b shows $\eta_{\rm res}^*$ plotted as a function of $\kap$ for several values of $t_c$.  The maximum efficiency is ${\eta_{\rm max}=98.4\%}$  and occurs at ${\kappa^*/2\pi=76}$ kHz and ${t_c/2\pi=0.5}$ GHz.  We note that this value of $\kap^*$ is well within the range of three-dimensional superconducting resonators which can have leakage rates as low as $\kap=0.1$ kHz \cite{reagorAPL13}.  \fig{fig:eff}b shows the  detector efficiency as a function of DQD bias in the optimal regime.

\section{Conclusions and Discussion} 
In summary, we have theoretically proposed and optimized a microwave photon detector based on a resonator-coupled double quantum dot, which could readily be integrated with current cQED technology. 
We show that very high quantum efficiency is possible with currently achievable values of the DQD-resonator coupling and DQD dissipation, and determined the parameter regime for near-unit efficiency. 
While we utilized purely charge states of the DQD in this work, our theoretical model can readily be applied to spin-charge hybridized DQD states, for example, the singlet and triplet states of Ref.~\cite{wongPRB15}, which are protected from charge noise dephasing and hence advantages for applications in quantum communication.

The proposed photon detector allows measurements of the input photon statistics as well, so that one could distinguish pure input Fock states from classical states by measuring the second order photon correlation \cite{wallsBook2012}.  However, determining the exact relation between photon and electric current noise is beyond the scope of this paper.  
To address this relation, one has to analyze the effect of fluctuations in the resonator photon number due to fluctuations in $\Ndot$, which results in $\s_z$ and current fluctuations and also in the backaction of the DQD on the resonator photons.
Note also that multi-photons states with $k$ photons in the resonator can also be detected by tuning the DQD to resonance with $k\w_{\rm in}$, but results in weaker signal since the multi-photon absorption transition amplitudes are reduced by the factor $g^k$ at weak coupling \cite{wielRMP02}.

\begin{acknowledgments}
We are thankful to  R. McDermott, B. Plourde, M. Schoendorf, F. Wilhelm, and C. Xu for fruitful discussions.
M.V. was supported by the Army Research Office under contract W911NF-14-1-0080 and NSF Grant No. DMR 0955500.  C.W. was supported by the Intelligence Community Postdoctoral Research Fellowship Program.
\end{acknowledgments}

\appendix 

\section{Charge relaxation in the DQD\label{sec:noise}}

Charge noise and phonons  couple to DQD charge density in an exact analogy with the DQD bias  and resonator field\cite{kulkarniPRB14}.  The associated dissipation thus is proportional to the square of the DQD charge dipole matrix element $\propto\sin^2\theta$.  Below we describe a minimal phenomenlogical model that incorporate these noise sources in the DQD dissipation.

\subsection{Relaxation rates due to phonons}

\begin{figure}[t]
\begin{center}
\includegraphics[width=0.7\linewidth]{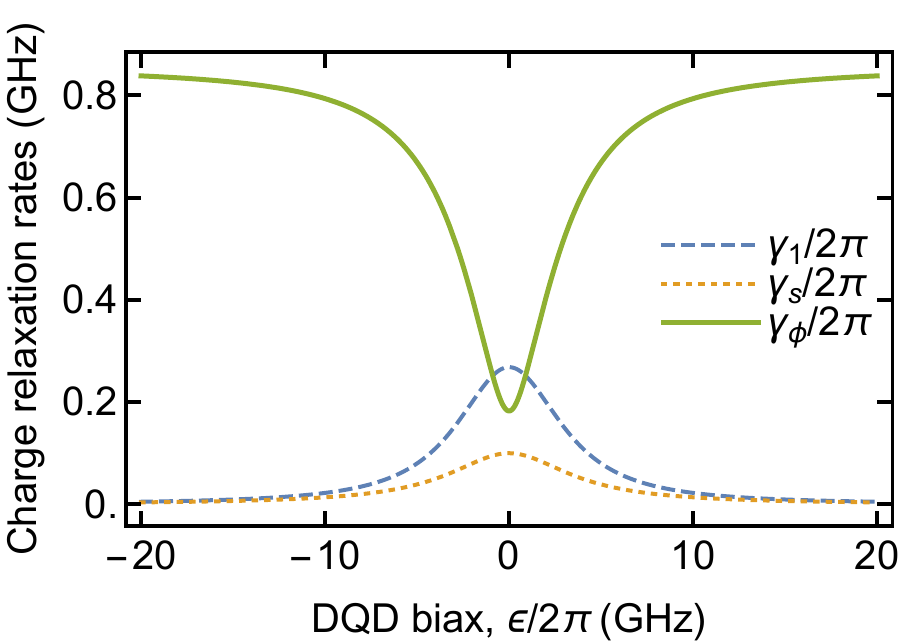}
\caption{The DQD charge relaxation rates, $T_1$ relaxation rate $\g_{1}$, difference in excitation and relaxation rates due to thermal phonons  $\g_s$, and quasistatic dephasing rate due to charge noise $\g_\phi$ as a function of the DQD bias $\e$.}
\label{fig:DQDrelax}
\end{center}
\end{figure}

The electron-phonon interaction Hamiltonian can be written as \cite{mahanBook}
\begin{align}
H_{\rm ph}&=\sin\theta\sum_{\mbk\lm}\beta_{\mbk\lm}(\s^\dag a_{\mbk\lm}+a_{\mbk\lm}^\dag\s)\,,
\label{eq:Hph}
\end{align}
where $\lm$ labels the phonon branches,
\[\beta_{\mbk\lm}=v_{\mbk\lm}\int d\mbr e^{i\mbk\cdot\mbr}\frac{|\phi_L(\mbr)|^2-|\phi_R(\mbr)|^2}{2}\]
where  $\phi_i(\mbr)$, are the localized (wannier) basis functions for the DQD electrons, and $v_{\mbk\lm}$ are coupling coefficients that depends on phonon pararmeters \cite{gullansPRL15}.

From Fermi's golden rule, the stimulated emission and absorption rates are given by
\begin{align}
\begin{pmatrix}
{\g^{\rm ph}_\up}\\{\g^{\rm ph}_\down}
\end{pmatrix}&=\g_s\begin{pmatrix}n_{\rm ph}(\w_\s)\\1+n_{\rm ph}(\w_\s)\end{pmatrix}
\end{align}
where $n_{\rm ph}(\w)=(e^{\hbar\w/k_BT}-1)^{-1}$ is the phonon thermal distribution.  The phonon-induced spontaneous emission rate is given by 
\[\g_s=\sin^2\theta J_{\rm ph}(\w_\s)~,\] where
\[J_{\rm ph}(\w)=2\pi\sum_{\mbk\lm}|\beta_\mbk|^2\de(\w-\w_{{\rm ph} \mbk\lm})\]
is the phonon spectral density and $\w_{{\rm ph} \mbk\lm}$ are the phonon dispersions. We assume a typical temperature of $50$ mK.  Note that{ phonons do not cause pure dephasing due to the vanishing phonon density of states at zero frequency \cite{huPRB12}.}

The phonon spectrum is material dependent. For silicon quantum dots, Ref.~\cite{wangPRL13} measures a charge relaxation times of up to $10$ ns at the DQD excitation frequency $\w_\s/2\pi=12$ GHz with the tunnel coupling $2t_c=5.9$ GHz.  We take a nominal value of $J_{\rm ph}(\w)=0.1$ GHz.

\subsection{Relaxation rates due to charge noise}

We assume a $1/f$ charge noise spectrum
\ben S_\e(\w)=\frac{c^2_\e}{|\w|}\,,\quad \w_l<|\w|<\w_h\,,\label{eq:epsnoise}\een
where  $c_\e=2~\mu {\rm eV}$ \cite{wongPRB16}.  
The depolarization rate is given by \cite{makhlinCP04}
\[\frac{\g^{\rm ch}_1(\e)}{2\pi}=\sin^2\theta\frac{S_\e(\w_\s)}{2\hbar^2}\,.\]
The low frequency part of the noise spectrum causes quasistatic fluctuations $\de\w_\s$ of the DQD excitation frequency.
The associated dephasing rate is given by \cite{wongPRB16}
\ben
\frac{\gamma _\phi}{2\pi }=\sqrt{\sum _{\alpha}\left(\frac{\p \w_\s}{\p\e}\frac{\e_{\rm rms}}{\sqrt{2}}\right)^2+\left(\frac{\p^2 \w_\s}{\p\e^2}\frac{ \e_{\rm rms}^2}{2}\right)^2}
\label{eq:gammaphi}
\een

The total charge relaxation rates $\g_1,\g_s,\g_\phi$ due to phonons and charge noise are plotted as a function of the DQD bias $\e$ in \fig{fig:DQDrelax}.
Charge noise induced pure dephasing, which is the dominant rate, is strongest at large $\e$ where the DQD eigenstates are pure charge states, and smallest at the charge transition $\e=0$, where the eigenstate are fully hybridized charge states, where the DQD energy is first order insensitive to bias noise.  The qualitative bias dependence of $\G_2$  follows that of $\g_\phi$, since the decrease in $\g_\phi$ is greater than  the increase in $\g_1$.  The bias dependence of $m_z^0$ and $\G_1$ follows that of $\g_1$ and $\g_s$.

\bibliography{physics}

\end{document}